\documentclass{iopart}
\usepackage{cite}
\usepackage{type1cm}
\usepackage[dvipdfm]{graphicx}
\usepackage{iopams}

\def\beq{\begin{equation}}
\def\eeq{\end{equation}}
\def\nbeq{\begin{equation*}}
\def\neeq{\end{equation*}}
\def\<{\langle}
\def\>{\rangle}
\def\rt#1{\sqrt{\mathstrut #1}}
\renewcommand{\d}{\partial}

\newtheorem{theo}{Theorem}
\newtheorem{corollary}{Corollary}

\makeatletter
\newcommand{\subscripts}[3]{%
  \@mathmeasure\z@\displaystyle{#2}%
  \global\setbox\@ne\vbox to\ht\z@{}\dp\@ne\dp\z@
  \setbox\tw@\box\@ne
  \@mathmeasure4\displaystyle{\copy\tw@_{#1}}%
  \@mathmeasure6\displaystyle{{#2}_{#3}}%
  \dimen@-\wd6 \advance\dimen@\wd4 \advance\dimen@\wd\z@
  \hbox to\dimen@{}\mathop{\kern-\dimen@\box4\box6}%
}
\makeatother

\begin{document}

\title{Phase transitions in systems with non-additive long-range interactions}
\author{Takashi Mori}
\address{
Department of Physics, Graduate School of Science,
The University of Tokyo, Bunkyo-ku, Tokyo 113-0033, Japan
}
\ead{mori@spin.phys.s.u-tokyo.ac.jp}

\begin{abstract}
We consider spin systems with long-range interactions in non-additive regime.
When the non-additive scaling limit is employed, the energy and the entropy compete and the system exhibits some phase transitions.
Such systems do not satisfy the additivity, which results in some unfamiliar properties related to phase transitions.
In this paper, the concept of additivity and its consequence are explained and the recent progress on statistical mechanics of long-range interacting systems are reviewed.
It is shown that the parameter space is clearly decomposed into the three regions according to the stability of the uniform state predicted by the mean-field theory.
Based on this parameter space decomposition, recent results on the exactness of  MF theory are explained.
When the interaction is non-negative (ferromagnetic), the analysis of the mean-field theory is exact 
and a typical spin configuration is always uniform in the canonical ensemble.
However, in the restricted canonical ensemble, i.e., the canonical ensemble with a restriction of the value of the magnetization, 
it is shown that the mean-field theory does not necessarily give the exact description of the system
and phase transitions between the mean-field uniform states (MF phase) and the inhomogeneous states (non-MF phase) occur.
A new finding is that when the interaction potential changes its sign depending on the distance, 
the non-MF phase appears even in the canonical ensemble.
\end{abstract}

\section{Introduction}

Additivity is a fundamental property of macroscopic systems.
Roughly speaking, additivity is defined as follows (we will give its precise definition later).
Let us consider a macroscopic system consisting of the two macroscopic subsystems $A$ and $B$.
When the energy of the total system is equal to the sum of the internal energies of $A$ and $B$ for any microscopic state,
this system is said to be additive.\footnote
{This condition is appropriate for a lattice system, but it is too strong in general.
An appropriate definition of the additivity is given in Sec.~\ref{sec:additivity}.}

In a short-range interacting system, it is expected that additivity holds 
because the interaction energy between the two subsystems is negligible compared to the bulk energy.
In contrast, additivity does not hold in a long-range interacting system~\cite{Campa_review2009,Les_Houches2009,Lecture_notes2002}.
When the two-body interaction potential decays slower than $1/r^d$, where $d$ is the spacial dimension and $r$ is the distance between two particles,
or when the interaction range is comparable with the system size,
the interaction energy between the two subsystems is not negligible compared to the internal energy of each subsystem,
and hence additivity does not hold.

When we consider statistical mechanics of such a system, some basic familiar properties derived from the assumption of additivity do not necessarily hold.
Usually additivity is taken for granted, but we should reconsider which properties are actually derived from additivity.

For example, the ensemble equivalence is a fundamental result of statistical mechanics~\cite{Ruelle_text}, but it relies on additivity.
To put it the other way around, in a long-range interacting system, two statistical ensembles, e.g., the microcanonical ensemble and the canonical ensemble,
can be inequivalent~\cite{Thirring1970,Hertel-Thirring1971,Ellis2000,Leyvraz2002,Bouchet2005}.
As a special consequence, the specific heat can be negative in the microcanonical ensemble while it is always non-negative in the canonical ensemble.

The other example, which is related to the ensemble equivalence, is the relation between the convexity or concavity of the thermodynamic function
and the thermodynamic stability~\cite{Landau_stat,Callen_text,Reichl_text}.
In the microcanonical ensemble, for instance, the entropy is always concave with respect to the energy in an additive system.
From the assumption of additivity, it is shown that if there were the energy interval where the entropy is not concave,
every state in this energy interval would be thermodynamically unstable; i.e., thermal fluctuation destroys such a state.
Because all the equilibrium states should be stable, the entropy should be a concave function of the energy.
However, when the system is non-additive, non-concavity of the entropy does not contradict the stability of equilibrium states.

When we apply the mean-field theory and calculate the thermodynamic function,
we sometimes encounter a non-convex or non-concave thermodynamic function.
In such a situation, we usually employ the Maxwell construction to make the thermodynamic function convex or concave~\cite{Callen_text,Reichl_text}.
This is an appropriate technique for an additive system.
Indeed, the mean-field theory with the Maxwell construction becomes exact in the limit 
such that the interaction range and the system size tend to infinity within the additive regime;
the interaction range is much shorter than the system size~\cite{Kac1963,van_Kampen1964,Lebowitz-Penrose1966}.
This limit is called the van der Waals limit and the above result is referred to as 
the Lebowitz-Penrose theorem~\cite{Lebowitz-Penrose1966,Lieb1966,Gates-Penrose1969,Gates-Penrose1970-1,Gates-Penrose1970-2}.

As was explained, the thermodynamic function may be non-convex or non-concave in a non-additive system.
Therefore, the Maxwell construction is not an appropriate procedure.
In this paper, we present the recent results on the extension of the relation between convexity or concavity of the thermodynamic function
and the thermodynamic stability into non-additive systems.
This extension also leads us to the extension of the Lebowitz-Penrose theorem into the non-additive regime.
The previous works~\cite{Mori2010_analysis,Mori2011_instability,Mori2012_microcanonical,Mori2012_equilibrium} assumed that
the two-body interaction is ferromagnetic for any distance.
In this paper, we will remove this assumption and discuss the possible consequences in Sec.~\ref{sec:non-negativity}.

The organization of the remaining part of this paper is following.
In Sec.~\ref{sec:preliminary}, we explain the setup and the notation.
In Sec.~\ref{sec:additivity}, the concepts of additivity and extensivity are explained in detail.
Some consequences of additivity are summarized in Sec.~\ref{sec:property}.
In Sec.~\ref{sec:MF}, we introduce the mean-field (MF) model, which is a very simple long-range interacting system.
In Sec.~\ref{sec:exactness}, we mention the central result (Theorem~\ref{theorem:main})
and the exactness of MF theory in the canonical ensemble is derived as a corollary of Theorem~\ref{theorem:main}.
It is rigorously shown that the system undergoes the phase transition between the MF phase and the non-MF phase in the restricted canonical ensemble.
In Sec.~\ref{sec:non-negativity}, we consider the case where the interaction potential is ferromagnetic in average but not fully ferromagnetic,
i.e., the condition C2 mentioned in Sec.~\ref{sec:preliminary} is removed.
There we show that the system undergoes the phase transition between the MF phase and the non-MF phase even in the canonical ensemble.
Section~\ref{sec:conclusion} concludes with a summary.

\section{Preliminary}
\label{sec:preliminary}

\subsection{Model}
We consider a classical spin system described by the following Hamiltonian:
\beq
H_h=-\frac{1}{2}\sum_{i,j}^N\gamma^d\phi(\gamma\bi{r}_{ij})\sigma_i\cdot\sigma_j-h\cdot\sum_{i=1}^N\sigma_i
\label{eq:Hamiltonian}
\eeq
Spins are on the $d$-dimensional cubic lattice.
The lattice distance is set to be unity.
The position of the site $i$ is denoted by $\bi{r}_i\in{\mathbb Z}^d\cap[-L/2,L/2)^d$, and the number of spins is given by $N=L^d$.
The $a$th component of $\bi{r}_i$ is denoted by $\bi{r}_i^{(a)}$, where $a=1,2,\dots,d$.
The distance between two sites $i$ and $j$ is denoted by $\bi{r}_{ij}$.
When the periodic boundary condition is imposed, this distance is interpreted as $\bi{r}_{ij}=(\bi{r}_i-\bi{r}_j)_{\rm P}$, where
\beq
(\bi{r}_i-\bi{r}_j)_{\rm P}^{(a)}:=
\cases{\bi{r}_i^{(a)}-\bi{r}_j^{(a)} & for $\left|\bi{r}_i^{(a)}-\bi{r}_j^{(a)}\right|\leq\displaystyle{\frac{L}{2}}$, \\
L-\left(\bi{r}_i^{(a)}-\bi{r}_j^{(a)}\right) & otherwise.}
\label{eq:PBC}
\eeq
It is pointed out that $\phi(\bi{x})$ is meaningful only for 
$$\bi{x}\in\left\{\bi{x}'-\bi{y}':\bi{x}',\bi{y}'\in\left[-\frac{\gamma L}{2},\frac{\gamma L}{2}\right)^d\right\}=:\Lambda_d.$$
Therefore, for $\bi{x}\notin\Lambda_d$, we may put $\phi(\bi{x})=0$.

Spin variables $\{\sigma_i\}$ and the uniform magnetic field $h$ are generally multi-component vectors.
$\sigma_i\cdot\sigma_j$ and $h\cdot\sigma_i$ should be interpreted as the inner products of two vectors.
We define ${\cal S}$ as the set of possible values of the spin variable.
It is arbitrary as long as the magnitude of the spin variable is finite; $|\sigma_i|<\infty$, $\forall\sigma_i\in{\cal S}$.
Some examples of ${\cal S}$ are listed in Table~\ref{table:spin_variable}.

\begin{table}[t]
\begin{center}
\begin{tabular}{l|l}
\hline
Ising model & ${\cal S}=\{+1,-1\}$ \\
3-state Potts model & ${\cal S}=\left\{\left(\begin{array}{c}1\\ 0\\ 0\end{array}\right), \left(\begin{array}{c}0\\ 1\\ 0\end{array}\right),
\left(\begin{array}{c}0\\ 0\\ 1\end{array}\right)\right\}$ \\
XY model & ${\cal S}=\left\{\left(\begin{array}{c}\cos\theta \\ \sin\theta\end{array}\right):\theta\in[0,2\pi)\right\}$ \\
(classical) Heisenberg model & ${\cal S}=\left\{
\left(\begin{array}{c}\sin\theta\cos\phi \\ \sin\theta\sin\phi \\ \cos\theta\end{array}\right):\theta\in[0,\pi],\phi\in[0,2\pi)\right\}$ \\
anisotropic Heisenberg model & ${\cal S}=\left\{
\left(\begin{array}{c}\rt{J_x}\sin\theta\cos\phi \\ \rt{J_y}\sin\theta\sin\phi \\ \rt{J_z}\cos\theta\end{array}\right)
:\theta\in[0,\pi],\phi\in[0,2\pi)\right\}$ \\
\hline 
\end{tabular}
\end{center}
\label{table:spin_variable}
\caption{Some examples of spin variables.}
\end{table}

The interaction potential $\phi(\bi{x})=\phi(-\bi{x})$ is assumed to satisfy the following conditions:
\begin{description}
\item[C1]: There exists a twice-differentiable, convex, and integrable function\footnote
{Integrable means $\displaystyle\int_0^{\infty}\psi(x)x^{d-1}dx<+\infty$ here.}
$\psi(x)$, defined in $x\in(0,\infty)$, such that $\forall\bi{x}\in\Lambda_d$,
$|\phi(\bi{x})|\leq\psi(x)$ and $|\nabla\phi(\bi{x})|\leq-d\psi(x)/dx=:-\psi'(x)$.
Here $x:=|\bi{x}|$ (in the periodic boundary condition, $x:=|\bi{x}_{\rm P}|$).
\item[C2 (non-negativity)]: $\phi(\bi{x})\geq 0$.
\end{description}
For some results stated in this paper, C2 is unnecessary, but in Sec.~\ref{sec:exactness_can}, we assume C2.
However, in Sec.~\ref{sec:decomposition} and Sec.~\ref{sec:non-negativity}, C2 is not assumed.

The small parameter $\gamma>0$ corresponds to the inverse of the interaction range~\cite{Kac1963}.
To consider long-range interacting systems, the limit of $\gamma\rightarrow 0$ is taken.
Here we consider the following two limits:
\begin{description}
\item[non-additive limit]: $L\rightarrow\infty$ with $\gamma L=1$ is fixed.
\item[van der Waals limit]: $\gamma\rightarrow 0$ after $L\rightarrow\infty$.
\end{description}
In the non-additive limit, the interaction range is comparable with the system size and such a system exhibits peculiar properties,
which will be explained in Sec.~\ref{sec:property}.
Essentially the same limit was explored in Refs.~\cite{Kiessling-Percus1995,Kiessling-Lebowitz1997}.
In the van der Waals limit, the interaction range is much longer than the lattice distance, but much shorter than the system size.
In that case, the system has additivity and it does not show the peculiarities.
It is well known that the MF theory with the Maxwell construction becomes exact in the van der Waals limit, 
which is referred to as the Lebowitz-Penrose theorem~\cite{Lebowitz-Penrose1966}.
As a result, thermodynamic functions are independent of the precise form of the interaction potential $\phi(\bi{x})$ in the van der Waals limit.
To understand what happens in the non-additive limit is the main subject of this work.

We now show some examples of interaction potentials $\phi(\bi{x})$.
In the non-additive limit, $\gamma L=1$, and hence
$\Lambda_d$ is a finite region. 
Therefore we can put $\phi(\bi{x})=0$ for all $\bi{x}\notin\Lambda_d$.
As a result, the condition C1 implies 
$$\int_{\mathbb{R}^d}\phi(\bi{x})d^dx=\int_{\Lambda_d}\phi(\bi{x})d^dx\leq\int_{\Lambda_d}\psi(x)d^dx<+\infty.$$
If we consider power-law interactions $\phi(\bi{x})\sim 1/x^{\alpha}$, the condition C1 requires $\alpha<d$.\footnote
{The case of $\phi(\bi{x})\sim 1/x^d$ for $x\ll 1$ (not for $x\gg 1$ because of the unusual scaling) is excluded in this setting, 
but we can treat it if we consider the following scaling of the Hamiltonian:
$$
H_h=-\frac{1}{2\ln L}\sum_{i,j}L^{-d}\phi(L^{-1}\bi{r}_{ij})\sigma_i\cdot\sigma_j-h\cdot\sum_i\sigma_i.
$$
}
We can also put $\phi(\bi{x})\sim e^{-x}$.
In this case, $\gamma^d\phi(\gamma\bi{r}_{ij})\sim (1/L)^d\exp[-\bi{r}_{ij}/L]$ 
and it expresses the exponentially decaying interaction with the interaction range $\xi\simeq L$.
It is also a sort of (non-additive) long-range interactions.

In the van der Waals limit, the condition C1 implies $\int_{{\mathbb R}^d}\phi(\bi{x})d^dx<\infty$
and the interactions which decay not faster than $1/r^d$ in long distance are excluded, which are treated within the non-additive limit. 

Hereafter, for simplicity, we use the same symbol ``${\rm Lim}$'' for both the non-additive limit and the van der Waals limit.
If this symbol is used in an equation, it implies that this equation holds both for the non-additive limit and for the van der Waals limit.

Finally, we should comment on the size-dependent scaling of the interaction in the non-additive limit.
If we write the interaction part as $-(1/2)\sum_{i,j}^NJ_{ij}\sigma_i\cdot\sigma_j$, 
$J_{ij}$ obeys the following scaling, $J_{ij}=(1/L^d)\phi(\bi{r}_{ij}/L)$ in the non-additive limit.
This size-dependent scaling of the interaction potential makes the system extensive (the definition of extensivity will be given in Sec.~\ref{sec:additivity})
and it is called ``Kac's prescription'' in the literatures~\cite{Kac1963}.
Kac's prescription should be viewed as a mathematical operation in order to extract the macroscopic properties of the system
in the regime where the energetic effect and the entropic effect compete with each other, i.e., in order to study phase transitions.

\subsection{Statistical ensemble}
We investigate the equilibrium properties of the model given by Eq.~(\ref{eq:Hamiltonian}) by the method of equilibrium statistical mechanics.
In this article, we mainly focus on the two statistical ensembles, the canonical ensemble and the restricted canonical ensemble.

In the canonical ensemble, each state is realized in the probability $e^{-\beta H_h}/\Xi(\beta,h)$, where $\Xi(\beta,h)$ is the partition function,
\beq
\Xi(\beta,h):=\sum_{\{\sigma_i\in{\cal S}\}}e^{-\beta H_h}.
\eeq
The free energy is defined by
\beq
G(\beta,h)=L^dg(\beta,h)=-\frac{1}{\beta}\ln\Xi(\beta,h).
\label{eq:G}
\eeq
The canonical ensemble describes the situation where the system is in contact with a thermal bath at the temperature $\beta^{-1}$.

In the restricted canonical ensemble, the value of the magnetization is restricted to $M=\sum_{i=1}^N\sigma_i$ and no magnetic field is applied.
Therefore, the probability of the state $\{\sigma_i\}$ is given by $e^{-\beta H_0}\chi(\sum_{i=1}^N\sigma_i=M)/Z(\beta,M)$,
where the partition function is given by
\beq
Z(\beta,m):=\sum_{\{\sigma_i\in{\cal S}\}}\chi\left(\sum_{i=1}^N\sigma_i=M\right)e^{-\beta H_0}.
\eeq
Here the Hamiltonian $H_0$ is given by putting $h=0$ in Eq.~(\ref{eq:Hamiltonian})
and the characteristic function $\chi(\cdot)$ is defined by $\chi({\rm True})=1$ and $\chi({\rm False})=0$.
The associated free energy is defined by
\beq
F(\beta,M)=L^df(\beta,m)=-\frac{1}{\beta}\ln Z(\beta,M),
\label{eq:F}
\eeq
where $m=M/L^d$ is the magnetization density.

Although we mainly focus on the above two types of canonical ensembles, sometimes we will give some comments on the microcanonical ensemble.
In the microcanonical ensemble, the energy of the system is held fixed, so that it describes the system isolated from environment.
The probability of the state $\{\sigma_i\}$ is given by $\chi(H_h\in(E-\delta E,E])/\Omega(E)$, where
\beq
\Omega(E,h):=\sum_{\{\sigma_i\in{\cal S}\}}\chi(H_h\in(E-\delta E,E]).
\eeq
The associated thermodynamic function is the entropy, which is defined by
\beq
S(E,h)=L^ds(\varepsilon,h):=\ln\Omega(E,h).
\eeq
It is straightforward to introduce the ``restricted microcanonical ensemble'', in which both the value of the energy and the value of the magnetization are held fixed.

\section{Additivity and extensivity from the statistical mechanical point of view}
\label{sec:additivity}

In this section, the concepts of additivity and extensivity are explained.
Let us consider the system described by the Hamiltonian~(\ref{eq:Hamiltonian}), which is in contact with a thermal bath at the temperature $\beta^{-1}$.
For simplicity, we put $h=0$ in this section since the external field plays a trivial role with regard to additivity and extensivity.
We imagine that the system is composed of two subsystems $A$ and $B$.
The Hamiltonian may be written as $H_{AB}=H_{A}+H_{B}+V$,
where $H_{X}=-(1/2)\sum_{i,j\in{X}}J_{ij}\sigma_i\cdot\sigma_j$ ($X$ is $A$ or $B$) and 
$V=-\sum_{i\in{A}, j\in{B}}J_{ij}\sigma_i\cdot\sigma_j$.
Here $J_{ij}=\gamma^{-d}\phi(\gamma\bi{r}_{ij})$.
When the magnetization is not conserved, equilibrium state is described by the canonical ensemble.

The concept of additivity is closely related to statistical independencce of these two macroscopic subsystems.
More precisely, the system is said to be additive if
\beq
P_{AB}(\beta,M_{A},M_{B})\approx P_{A}(\beta,M_{A})P_{B}(\beta,M_{B}),
\label{eq:additivity}
\eeq
for any pair of values of $(M_{A},M_{B})$.
Here, $P_{AB}(\beta,M_{A},M_{B})$ denotes the probability that the magnetizations of $A$ and $B$ are equal to $M_{A}$ and $M_{B}$
in the system described by $H_{AB}$.
Similarly, $P_{X}(\beta,M_{X})$ denotes the probability that the magnetization of the system $X$ is equal to $M_{X}$
{\it when the system described by $H_{X}$ alone is in contact with a thermal bath}.
It is noted that $P_{A}(\beta,M_{A})\neq\sum_{M_{B}}P_{AB}(\beta,M_{A},M_{B})$, in general.
If Eq.~(\ref{eq:additivity}) is satisfied, the statistical property of one of the macroscopic subsystems is not influenced by the presence of the others.

According to equilibrium statistical mechanics, such probabilities are related to the free energy with some constraints.
If $F_{AB}(\beta,M_{A},M_{B})$ is defined as the free energy associated with $H_{AB}$
under the constraint that the magnetizations of $A$ and $B$ are equal to $M_{A}$ and $M_{B}$, respectively, that is,
\begin{eqnarray}
F_{AB}(\beta,M_{A},M_{B})=-\frac{1}{\beta}\ln Z(\beta,M_{A},M_{B})
\nonumber \\
=-\frac{1}{\beta}\ln\sum_{\{\sigma_i\in{\cal S}\}}e^{-\beta H_0}
\chi\left(\sum_{i\in{A}}\sigma_i=M_{A}\cap\sum_{i\in{B}}\sigma_i=M_{B}\right),
\end{eqnarray}
we can express $P_{AB}(\beta,M_{A},M_{B})$ in terms of free energies as follows:
\beq
P_{AB}(\beta,M_{A},M_{B})=\exp\left\{\beta\left[G_{AB}(\beta,0)-F_{AB}(\beta,M_{A},M_{B})\right]\right\}.
\label{eq:P_AB}
\eeq
The free energy $G$ is defined in Eq.~(\ref{eq:G}).
In the similar way, $P_{X}(\beta,M_{X})$ is expressed as
\beq
P_{X}(\beta,M_{X})=\exp\left\{\beta\left[G_{X}(\beta,0)-F_{X}(\beta,M_{X})\right]\right\}.
\label{eq:P_X}
\eeq

From Eqs.~(\ref{eq:P_AB}) and (\ref{eq:P_X}), the condition of additivity, Eq.~(\ref{eq:additivity}), is rewritten as
\begin{eqnarray}
F_{AB}(\beta,M_{A},M_{B})-F_{A}(\beta,M_{A})-F_{B}(\beta,M_{B})
\nonumber \\
=G_{AB}(\beta,0)-G_{A}(\beta,0)-G_{B}(\beta,0).
\label{eq:add_free}
\end{eqnarray}
If the above equality holds for any $(M_{A},M_{B})$, the system is said to be {\it additive}.
In addition, if the RHS of Eq.~(\ref{eq:add_free}) is zero, the system is said to be {\it extensive}.

Usually, in a short-range interacting system, both the LHS and the RHS of Eq.~(\ref{eq:add_free}) are equal to zero
because the interaction energy between the two subsystem is of the order of the surface area rather than the volume.
Thus many short-range interacting systems are extensive and additive.

On the other hand, in long-range interacting systems, extensivity and additivity are not satisfied, in general.
However, the lack of extensivity is apparent;
By applying Kac's prescription, that is, by putting $J_{ij}=(1/L^d)\phi(\bi{r}_{ij}/L)$, we can make the system extensive
since the energy per spin is held fixed.
The important point is that even if we apply Kac's prescription, the system is still non-additive.
Thus non-additivity is an essential feature of long-range interacting systems.

\section{Some consequences of additivity and non-additivity}
\label{sec:property}

Some familiar properties of macroscopic systems are actually resulted from additivity.
If the system is non-additive, those properties are no longer a matter of course.
We have to reconsider the basic properties of the system.
From now on, we review some of the properties, in which additivity plays an important role.

\subsection{Susceptibilities in the restricted canonical ensemble}

The susceptibility is defined as the response of the magnetization with respect to the small change of the magnetic field, $\chi:=\d M/\d h$.
In the canonical ensemble, $M=-\d G/\d h$, and hence $\chi=-\d^2G/\d h^2$.
In the restricted canonical ensemble, $h=\d F/\d M$, and hence $\chi=(\d^2F/\d M^2)^{-1}$.
The meaning of the magnetic field $h$ in the restricted canonical ensemble is the following;
If the additional spin $\sigma'$ is attached into the system very weakly, the probability of $\sigma'$ is given by
$P(\sigma')=e^{\beta h\cdot\sigma'}/\sum_{\sigma'\in{\cal S}}e^{\beta h\cdot\sigma'}$.

Let us consider the restricted canonical ensemble.
We virtually divide the system into identical subsystems $A$ and $B$.
If the values of magnetization of $A$ and $B$ are equal to $M_{A}$ and $M_{B}$, respectively,
due to additivity, the probability of this state is proportional to
$\exp[-\beta(F_{A}(\beta,M_{A})+F_{B}(\beta,M_{B}))]$.
When the system is in equilibrium, $M_{A}=M_{B}=M^{\rm eq}$, 
where $M^{\rm eq}$ is the equilibrium value of the magnetization.
The equilibrium state corresponds to the most probable state, 
so that the free energy $F_{A}(\beta,M_{A})+F_{B}(\beta,M_{B})$ should take the maximum value at $M_{A}=M_{B}=M^{\rm eq}$.
Let us imagine that the magnetization fluctuates at some instantaneous time 
and $(M_{A},M_{B})$ becomes $(M^{\rm eq}+\delta M,M^{\rm eq}-\delta M)$.
Such a fluctuation occurs at the probability proportional to $\exp[-\beta(F_{A}(\beta,M^{\rm eq}+\delta M)+F_{B}(\beta,M^{\rm eq}-\delta M))]$.
By expanding with respect to $\delta M$, we have
\begin{eqnarray}
\fl F_{A}(\beta,M^{\rm eq}+\delta M)+F_{B}(\beta,M^{\rm eq}-\delta M)
\nonumber \\ \fl
=F_{A}(\beta,M^{\rm eq})+F_{B}(\beta,M^{\rm eq})
+\delta M(h_{A}-h_{B})+\frac{(\delta M)^2}{2}(\chi_{A}^{-1}+\chi_{B}^{-1}).
\label{eq:expansion}
\end{eqnarray}
Since the two subsystems are assumed to be identical, $h_{A}=h_{B}$ and $\chi_{A}=\chi_{B}$ in equilibrium.
Therefore, in the restricted canonical ensemble, the susceptibility must be non-negative $\chi\geq 0$ 
so that the equilibrium state is the most probable state.

When the system is not additive, the above reasoning does not work and the susceptibility can be negative in the restricted canonical ensemble.
It is noted that even if the system is non-additive, the susceptibility is non-negative in the canonical ensemble.
It follows from another expression of the susceptibility,
\beq
\chi=-\frac{\d^2G}{\d h^2}=\beta(\< M^2\>-\< M\>^2),
\label{eq:chi_c}
\eeq
where $\<{\cal O}\>:=\sum_{\{\sigma_i\in{\cal S}\}}{\cal O}e^{-\beta H_{\rm h}}/\Xi(\beta,h)$.
Mathematically the RHS of Eq.~(\ref{eq:chi_c}) is always non-negative for $\beta>0$.
Therefore, when the susceptibility takes the negative value in the restricted canonical ensemble,
it implies that the canonical ensemble is not thermodynamically equivalent to the restricted canonical ensemble.
This feature is called {\it ensemble inequivalence}.

Similarly, the specific heat must be non-negative in an additive system.
The specific heat is always non-negative in the canonical ensemble and the restricted canonical ensemble, regardless of whether the system is additive or non-additive.
However, it can be negative in the microcanonical ensemble for a non-additive system.
Negative specific heats have been experimentally observed in small systems~\cite{d'Agostino2000,Schmidt2001,Gobet2002}.

\subsection{Convexity of the free energy and thermodynamic stability}

In an additive system, the free energy density $f(\beta,m)$ should be a convex function of $m$, which is implied by the non-negative susceptibility.
It is recognized as follows.
Let us consider a system in the restricted canonical ensemble.
We choose some $m_1$, $m_2$, and some $\lambda\in[0,1]$.
The system is assumed to be composed of the two subsystems, one of which has $\lambda N$ spins and the magnetization density $m_1$
and the other of which has $(1-\lambda)N$ spins and the magnetization density $m_2$.
Due to the additivity, the free energy density of this state is given by $\lambda f(\beta,m_1)+(1-\lambda)f(\beta,m_2)$.
Since the equilibrium state corresponds to the most probable state under a given restriction,
the equilibrium free energy must satisfy
\beq
f(\beta,\lambda m_1+(1-\lambda)m_2)\leq \lambda f(\beta,m_1)+(1-\lambda)f(\beta,m_2)
\eeq
for any $m_1$, $m_2$, and $\lambda\in[0,1]$.
It means that the free energy is convex in an additive system.

However, some approximations lead us to a non-convex free energy.
For example, the MF theory assumes that the system is homogeneous, which sometimes causes non-convexity of the free energy.
Such a non-convex free energy is denoted by $\tilde{f}(\beta,m)$.

\begin{figure}[t]
\begin{center}
\includegraphics[clip,width=4cm]{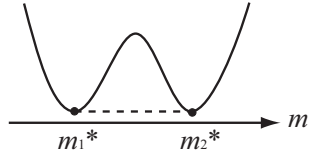}
\caption{Non-convex free energy $\tilde{f}(\beta,m)$. The dashed line is its convex envelope $\tilde{f}^{**}(\beta,m)$.}
\label{fig:separation}
\end{center}
\end{figure}

If $m$ is in the non-convex region, i.e., $m\in(m_1^*,m_2^*)$ in Fig.~\ref{fig:separation}, there are some $m_1$, $m_2$, and $\lambda\in(0,1)$ such that
$m=\lambda m_1+(1-\lambda)m_2$ and $\tilde{f}(\beta,m)\geq\lambda\tilde{f}(\beta,m_1)+(1-\lambda)\tilde{f}(\beta,m_2)$.
It implies that the assumption of homogeneity is broken down and the phase separation occurs.
If we allow the phase separation, the least free energy is realized for $m_1=m_1^*$ and $m_2=m_2^*$,
where $m_1^*$ and $m_2^*$ are the double tangent points, 
and the true free energy is on the double tangent line of $\tilde{f}(\beta,m)$; see Fig.~\ref{fig:separation}.
This double tangent line procedure is called the {\it Maxwell construction}.

In this way, in an additive system, the convexity of the free energy is ensured by the phase separation,
which typically occurs when the system undergoes a first order phase transition.
The Maxwell construction makes the free energy convex, $\tilde{f}(\beta,m)\rightarrow \tilde{f}^{**}(\beta,m)$,
where $f^{**}(\beta,m)$ is the {\it convex envelope} of $f(\beta,m)$, i.e. the maximum convex function not exceeding $f(\beta,m)$.

The condition of the phase coexistence is that $h=\d f/\d m$ is identical at $m=m_1^*$ and $m=m_2^*$, which is indicated by the double tangent line procedure
and also by the discussion in the previous subsection, see Eq.~(\ref{eq:expansion}).
Therefore, at a first order transition point, the magnetic field should be continuous if the system is additive.
Similarly, in the microcanonical ensemble for an additive system, the entropy is a concave function of the energy
and the temperature should be continuous at the first order transition point, as long as several phases can coexist there.

In a non-additive system, the situation becomes different.
Due to the large interaction between the two subsystems, even if we consider a state with two distinct phases with the magnetizations $m_1^*$ and $m_2^*$,
the free energy is not on the double tangent line, $f(\beta,m_1^*,m_2^*)\neq \lambda f(\beta,m_1^*)+(1-\lambda) f(\beta,m_2^*)$.
If the energy gain due to the phase separation is very large, 
the free energy in a state of the phase coexistence is larger than the free energy in a homogeneous state.
In this case, the solid line can give the correct free energy and the negative susceptibility does not contradict the thermodynamic stability in the above sense.

\subsection{Ensemble equivalence and inequivalence}

We discuss the relation between the canonical ensemble and the restricted canonical ensemble.
The partition function $\Xi(\beta,h)$ is expressed as
\begin{eqnarray}
\Xi(\beta,h)&=\sum_{\{\sigma_i\in{\cal S}\}}e^{-\beta H_h}
\nonumber \\
&=\sum_{\{\sigma_i\in{\cal S}\}}\sum_M\chi\left(\sum_{i=1}^N\sigma_i=M\right)e^{-\beta(H_0-h\cdot M)}
\nonumber \\
&=\sum_Me^{\beta h\cdot M}\sum_{\{\sigma_i\in{\cal S}\}}\chi\left(\sum_{i=1}^N\sigma_i=M\right)e^{-\beta H_0}
\nonumber \\
&=\sum_Me^{-\beta N[f(\beta,m)-h\cdot m]}
\nonumber \\
&\approx \exp\left[-N\beta\inf_m(f(\beta,m)-h\cdot m)\right].
\end{eqnarray}
The free energy density $g(\beta,h)$ is given by
\beq
g(\beta,h)=\inf_m\left[f(\beta,m)-h\cdot m\right].
\label{eq:g_from_f}
\eeq
This formula is always correct.
The free energy $g(\beta,h)$ is calculated from the Legendre transformation of $f(\beta,m)$.

However, it is not always possible to know $f(\beta,m)$ from $g(\beta,h)$.
The ``inverse'' transformation
\beq
f^{**}(\beta,m)=\sup_h\left[g(\beta,h)+h\cdot m\right]
\label{eq:f_from_g}
\eeq
yields only $f^{**}$, the convex envelope of $f(\beta,m)$.
The proof of Eq.~(\ref{eq:f_from_g}) is given in \ref{app:convex_envelope}.
If the free energy $f(\beta,m)$ is convex with respect to $m$, which is always the case in an additive system, $f(\beta,m)=f^{**}(\beta,m)$,
and thus
\beq
f(\beta,m)=\sup_h\left[g(\beta,h)+h\cdot m\right] \quad \textrm{if $f(\beta,m)$ is convex}.
\label{eq:f_from_g_additive}
\eeq
Equations~(\ref{eq:g_from_f}) and (\ref{eq:f_from_g_additive}) 
mean that the canonical ensemble is thermodynamically equivalent to the restricted canonical ensemble.

In a non-additive system, the free energy in the restricted canonical ensemble is not necessarily convex, as we have seen in the previous subsection.
In this case, we cannot know $f(\beta,m)$ from the knowledge of $g(\beta,h)$ and thermodynamic predictions made by the two ensembles may be different.
This is called {\it ensemble inequivalence}~\cite{Thirring1970,Hertel-Thirring1971,Ellis2000,Barre2001}
; for an introductory review, see Ref.~\cite{Touchette_review2004}.
Historically, the ensemble inequivalence was found in cosmology; see Ref.~\cite{Chavanis_review2006} for a review.

\section{Mean-field models}
\label{sec:MF}

Long-range interacting systems are representative of realistic non-additive systems.
They are described by the Hamiltonian~(\ref{eq:Hamiltonian}), but it seems to be difficult to solve the problem for a realistic interaction potential.
If $\phi(\bi{x})$ is a constant, $\phi(\bi{x})=1$, spacial geometry plays no role and it is expected to be able to calculate the free energy explicitly.

Putting $\phi(\bi{x})=1$, the Hamiltonian becomes
\beq
H_h^{\rm (MF)}=-\frac{1}{2N}\sum_{i,j}^N\sigma_i\cdot\sigma_j-h\cdot\sum_{i=1}^N\sigma_i,
\eeq
which is referred to as the MF model.
In the MF model, the Hamiltonian depends only on the magnetization,
\beq
H_h^{\rm (MF)}=N\left(-\frac{1}{2}m^2-h\cdot m\right),
\eeq
which makes the analysis tractable.
If we define the {\it spin configurational entropy} (or simply configurational entropy) as
\beq
\fl
s(m):=\lim_{N\rightarrow\infty}\frac{1}{N}\ln\left(\textrm{\# of states } \{\sigma_i\in{\cal S}\}_{i=1}^N 
\textrm{ with } \frac{1}{N}\sum_{i=1}^N\sigma_i=m\right),
\label{eq:config_entropy}
\eeq
the MF free energy in the restricted canonical ensemble is generally written as
\beq
f_{\rm MF}(\beta,m)=-\frac{1}{2}m^2-\frac{1}{\beta}s(m).
\label{eq:MF_free}
\eeq

Here let us explicitly calculate the MF free energy $f_{\rm MF}(\beta,m)$ for the Ising variable, ${\cal S}=\{+1,-1\}$.
Since the number of states for a given magnetization density is $N!/[((N+M)/2)!((N-M)/2)!]$,
\begin{eqnarray}
f_{\rm MF}(\beta,m)=-\frac{1}{N\beta}\ln\left[\frac{N!}{\left(\frac{N+M}{2}\right)!\left(\frac{N-M}{2}\right)!}e^{-\beta H_0^{\rm (MF)}}\right]
\nonumber \\
\approx -\frac{1}{2}m^2+\frac{1}{\beta}\left(\frac{1+m}{2}\ln\frac{1+m}{2}+\frac{1-m}{2}\ln\frac{1-m}{2}\right),
\end{eqnarray}
where we used Stirling's formula.
This free energy is convex for $\beta\leq 1$ and becomes non-convex for $\beta>1$.
We define the spinodal magnetization $m_{\rm sp}(\beta)>0$ by $\left.\d^2f_{\rm MF}(\beta,m)/\d m^2\right|_{m=m_{\rm sp}(\beta)}=0$ 
and $m_{\rm eq}(\beta)$ as a non-negative value of $m$ for which $f_{\rm MF}(\beta,m)$ takes the minimum value.
When $|m|>m_{\rm sp}(\beta)$, the susceptibility is positive.
On the other hand, when $|m|<m_{\rm  sp}(\beta)$, it is negative.
The quantity $m_{\rm eq}(\beta)$ corresponds to the equilibrium magnetization at the inverse temperature $\beta$ and $h=0$ in the canonical ensemble.

When $h=0$, the MF model undergoes the second order phase transition at $\beta=1$ in the canonical ensemble.
On the other hand, in the restricted canonical ensemble, there is no phase transition since $f_{\rm MF}(\beta,m)$ is analytic.
In this way, due to the ensemble inequivalence, the nature of the phase transition is totally different according to the specific ensemble.

Since the MF models are tractable analytically and numerically, 
the MF models have been studied extensively to understand the universal properties of long-range interacting systems.
Detailed nature of ensemble inequivalence has been clarified by analyzing several MF models~\cite{Torcini-Antoni1999,Barre2001,Ellis2004,Campa2007,Kastner2010}.
Not only their equilibrium properties, but also their dynamical aspects have triggered interests in recent years.
Mean-field models show interesting dynamical properties such as 
the existence of quasi-stationary states (or prethermalization) and associating phase transitions~\cite{Antoni-Ruffo1995,Antoniazzi2007,Baldovin2009,Gupta-Mukamel2010,Kastner2011,Sciolla-Biroli2011,Benetti2012}
and the ergodicity breaking in a finite system~\cite{Mukamel2005}.

However, it has to be said that the MF models belong to just a special class of long-range interacting systems 
in the sense that the MF models are not able to capture the spacial structure due to the long-range nature of an interaction.
Such spacial structure is expected to be widely observed in long-range interacting systems.
Therefore, it is important to understand the non-additive limit of the system 
whose potential $\phi(\bi{x})$ is not a constant~\cite{Campa2000,Cannas2000,Tamarit2000,Barre2005}.

\subsection*{Mean-field models and the van der Waals limit with the Maxwell construction}

The analysis of the MF model (the most naive version of the MF theory) also helps us to understand the thermodynamic behavior of an {\it additive} system.
In an additive system, the free energy should be convex for any temperature, but the MF free energy is not convex in general.
This discrepancy is avoided by considering the phase separation.
As was discussed in the previous subsection, the phase separation makes the free energy convex, which corresponds to the Maxwell construction.
By applying the Maxwell construction, we obtain a convex free energy $f_{\rm MF}^{**}(\beta,m)$.
The analysis of the MF model with the Maxwell construction is expected to well describe a system which is additive but whose interaction range is relatively long.
Indeed, Lebowitz-Penrose theorem states that the free energy density of a system in the {\it van der Waals limit} is exactly equal to $f_{\rm MF}^{**}(\beta,m)$
under the normalization $\int_{{\mathbb R}^d}d^dx\phi(\bi{x})=1$.

In the van der Waals limit, $f_{\rm MF}(\beta,m)$ is regarded as the free energy associated with a state with a homogeneous magnetization profile,
and while $f_{\rm MF}^{**}(\beta,m)$ is regarded as the free energy associated with a state of the phase coexistence.
The sign of the susceptibility determines the stability of a given state
and the parameter space can be decomposed into the following three regions:
\begin{description}
\item[]{\it Parameter space decomposition in the van der Waals limit.}
\item[A]: $|m|\geq m_{\rm eq}(\beta)$. In this region, a homogeneous state is globally stable.
\item[B]: $m_{\rm sp}(\beta)<|m|<m_{\rm eq}(\beta)$. In this region, a homogeneous state is locally stable but globally unstable.
\item[C]: $|m|\leq m_{\rm sp}(\beta)$. In this region, a homogeneous state is locally unstable.
\end{description}

In the canonical ensemble, $g_{\rm MF}(\beta,h)$ is computed from Eq.~(\ref{eq:g_from_f}).
The probability of the value of the magnetization being $m$ in equilibrium is proportional to $\exp[-N\beta(f_{\rm MF}(\beta,m)-h\cdot m)]$.
Therefore, a local minimum point of $f_{\rm MF}(\beta,m)-h\cdot m$ is considered to represent a metastable state.
In the MF model, the lifetime of a metastable state diverges in the thermodynamic limit, because the free energy barrier is proportional to the number of spins.
As a result, we can precisely define the metastability for the MF models in the van der Waals limit.~\cite{Penrose-Lebowitz1971}

\section{Parameter space decomposition and exactness of mean-field theory}
\label{sec:exactness}

As was mentioned in Sec.~\ref{sec:preliminary} and explained in Sec.~\ref{sec:MF}, 
in the van der Waals limit, the thermodynamic functions are independent of $\phi(\bi{x})$
and they are given by the MF theory with the Maxwell construction (Lebowitz-Penrose theorem~\cite{Lebowitz-Penrose1966}).
In the non-additive limit, it has been also indicated that thermodynamic functions do not depend on the interaction potential 
and are given by the MF ones but {\it without} the Maxwell construction.
It is called {\it exactness of mean-field theory}~\cite{Cannas2000}.
Exactness of MF theory can be viewed as a generalization of the Lebowitz-Penrose theorem to the non-additive limit.

The recent works~\cite{Mori2010_analysis,Mori2011_instability,Mori2012_microcanonical,Mori2012_equilibrium} have rigorously shown that
when the interaction potential satisfies the conditions C1 and C2,
the MF theory is always exact in the canonical ensemble 
but it is not in the restricted canonical ensemble and in the microcanonical ensemble~\cite{Mori2012_microcanonical}.
This result is also proven for quantum spin systems~\cite{Mori2012_equilibrium}.
In this section, the result on exactness of MF theory is reviewed for classical spins described by the Hamiltonian~(\ref{eq:Hamiltonian})
in the canonical and restricted canonical ensembles.

In this section we repeat essentially the same result given in Refs.~\cite{Mori2010_analysis,Mori2011_instability},
but here I present the result in a different way.
The result for the non-additive limit is compared with the well known result for the van der Waals limit.
It will help us to understand the result more deeply.
Especially, it is emphasized that the result given in this section is a generalization of the parameter space decomposition
in the van der Waals limit given in Sec.~\ref{sec:MF}.

\subsection{Parameter space decomposition in the non-additive limit}
\label{sec:decomposition}

In this subsection, we mention the central result, i.e., the parameter space decomposition in the non-additive limit.
The results in subsections~\ref{sec:exactness}, \ref{sec:non-MF}, and \ref{sec:non-negativity} are derived by this parameter space decomposition.
We first mention the central result as a theorem, and then explain its implications
in later subsections~\ref{sec:exactness}, \ref{sec:non-MF}, and \ref{sec:non-negativity}.
Finally, the strategy of the proof is sketched in Sec.\ref{sec:proof}.
We assume the periodic boundary condition in this section.

Before mentioning the result, some quantities need to be introduced.
In the periodic boundary condition, $\phi(\bi{x})$ is expressed by the Fourier coefficients $\{\phi_{\bi{n}}\}_{\bi{n}\in{\mathbb Z}^d}$ as
\beq
\phi(\bi{x}_{\rm P})=\sum_{\bi{n}\in{\mathbb Z}^d}\phi_{\bi{n}}e^{2\pi i\bi{n}\cdot\bi{x}}
=\sum_{\bi{n}\in{\mathbb Z}^d}\phi_{\bi{n}}\cos(2\pi\bi{n}\cdot\bi{x}),
\eeq
where we used $\phi(\bi{x})=\phi(-\bi{x})$.
The inverse transformation is
\beq
\phi_{\bi{n}}=\int_{\Omega_1^d}d^dx\phi(\bi{x}_{\rm P})e^{-2\pi i\bi{n}\cdot\bi{x}}
=\int_{\Omega_1^d}d^dx\phi(\bi{x})\cos(2\pi\bi{n}\cdot\bi{x}).
\eeq
Here $\Omega_1^d:=[-1/2,1/2)^d$ denotes the $d$-dimensional unit hypercube.
Since $\phi(\bi{x})$ is real, $\phi_{\bi{n}}\in{\mathbb R}$.
If $\phi(\bi{x})$ satisfies C2, $\phi_{\bi{n}}\leq\phi_0$ for any $\bi{n}\in{\mathbb Z}^d$.
This property is essential for the results in subsections~\ref{sec:exactness}.
We define $\phi_{\rm max}$ as
\beq
\phi_{\rm max}:=\max_{\bi{n}\in{\mathbb Z}^d\backslash 0}\phi_{\bi{n}}.
\label{eq:phi_max}
\eeq
When C2 is satisfied, $\phi_{\rm max}\leq\phi_0$.
Hereafter, we normalize the interaction potential as
\beq
\phi_0=\int_{\Omega_1^d}d^dx\phi(\bi{x})=1.
\eeq

We define the matrix ${\sf H}(\beta,m)$, whose component ${\sf H}_{ab}(\beta,m)$ is given by
\beq
{\sf H}_{ab}(\beta,m):=\frac{\d^2f_{\rm MF}}{\d m^{(a)}m^{(b)}}(\beta,m).
\eeq
The $a$-th component of $m$ is denoted by $m^{(a)}$ (Remember that a spin variable $\sigma_i$ is generally a multi-component quantity).
When the spin variable is scalar like in the Ising model, this matrix is simply ${\sf H}(\beta,m)=\d^2f_{\rm MF}(\beta,m)/\d m^2$.
Further we define $\lambda(\beta,m)$ as the smallest eigenvalue of ${\sf H}(\beta,m)$.

Now we present the central result, that is, the parameter space decomposition in the non-additive limit,
which is a generalization of that in the van der Waals limit given in Sec.~\ref{sec:MF}.
\begin{theo}[Parameter space decomposition in the non-additive limit]\label{theorem:main}
We consider the Hamiltonian~(\ref{eq:Hamiltonian}) on the $d$-dimensional regular lattice with the periodic boundary condition.
If the interaction potential satisfies C1 and is normalized as $\int_{\Omega_1^d}d^dx\phi(\bi{x})=1$, 
the parameter space $(\beta,m)$ is decomposed into the following three regions in the non-additive limit:
\begin{description}
\item[A]: $\left\{(\beta,m)|f_{\rm MF}(\beta\phi_{\rm max},m)=f_{\rm MF}^{**}(\beta\phi_{\rm max},m)\right\}$.
In this region, a typical equilibrium magnetization profile is homogeneous and $f(\beta,m)=f_{\rm MF}(\beta,m)$.
\item[B]: $\left\{(\beta,m)|f_{\rm MF}(\beta\phi_{\rm max},m)>f_{\rm MF}^{**}(\beta\phi_{\rm max},m)\textrm{ and }
\lambda(\beta\phi_{\rm max},m)\geq 0\right\}$.
In this region, a homogeneous state predicted by the MF theory is locally stable.
If this state is globally stable, $f(\beta,m)=f_{\rm MF}(\beta,m)$, otherwise 
$$f_{\rm MF}(\beta,m)-\phi_{\rm max}\Delta f_{\rm MF}(\beta\phi_{\rm max})\leq f(\beta,m)<f_{\rm MF}(\beta,m),$$
where $\Delta f_{\rm MF}:=f_{\rm MF}-f_{\rm MF}^{**}$.
\item[C]: $\left\{(\beta,m)|\lambda(\beta\phi_{\rm max},m)<0\right\}$.
In this region, a homogeneous state predicted by the MF theory is locally unstable,
and thus a typical equilibrium configuration is inhomogeneous.
The free energy satisfies the inequality
$$f_{\rm MF}(\beta,m)-\phi_{\rm max}\Delta f_{\rm MF}(\beta\phi_{\rm max})\leq f(\beta,m)<f_{\rm MF}(\beta,m).$$
\end{description}
\end{theo}

For the precise meaning of the term ``stable'' or ``unstable'', see Sec.~\ref{sec:proof}.
This result can be viewed as a generalization of the parameter space decomposition in the van der Waals limit.
Indeed, if we put $\phi_{\rm max}=1$, we recover the parameter space decomposition in the van der Waals limit.
As we will see later, the van der Waals limit corresponds to formally putting $\phi(\bi{x})=\delta(\bi{x})$ in the non-additive limit
when the condition C2 is satisfied.
Therefore, $\phi_{\bi{n}}=1$ for any $\bi{n}\in{\mathbb Z}^d$ in the van der Waals limit.

For simplicity, let us consider the case of one-component spin variable.
In that case, $\lambda(\beta\phi_{\rm max},m)=\d^2f_{\rm MF}(\beta\phi_{\rm max})/\d m^2$. 
In the van der Waals limit, the negative susceptibility, $\d^2f_{\rm MF}(\beta,m)/\d m^2<0$, 
implies local instability of the homogeneous state predicted by the MF theory.
On the other hand, in the non-additive limit, the negative susceptibility does not automatically mean instability as we have seen in Sec.~\ref{sec:additivity}.
The parameter space decomposition implies that the susceptibility at the potential-dependent {\it fictitious} temperature $(\beta\phi_{\rm max})^{-1}$
determines the stability.

\begin{figure}[t]
\begin{center}
\includegraphics[clip,width=8cm]{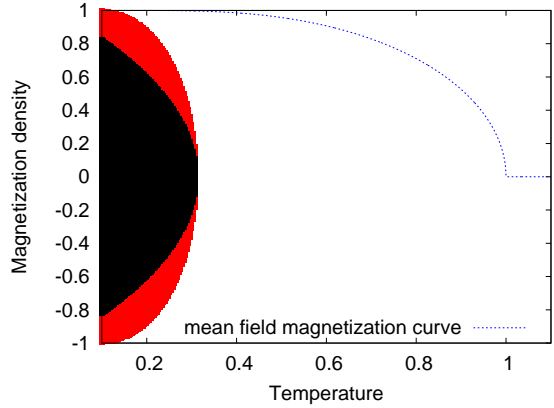}
\caption{The phase diagram on the $(\beta^{-1},m)$ plane when $\phi_{\rm max}=0.31$.
The white, red (gray), and the black regions correspond to the region A, the region B, and the region C, respectively.
The dashed line is the mean-field magnetization curve at $h=0$, which is fully in the region A.}
\label{fig:phase_TM}
\end{center}
\end{figure}

As an example, let us consider the case of the Ising variables, ${\cal S}=\{+1,-1\}$.
The condition $f_{\rm MF}(\beta,m)=f_{\rm MF}^{**}(\beta,m)$ is identical to $|m|\geq m_{\rm eq}(\beta)$.
The region of $\d^2f_{\rm MF}(\beta,m)/\d m^2\geq 0$ corresponds to $|m|\geq m_{\rm sp}(\beta)$.
Therefore, we can also express the three parameter regions A, B, and C as follows:
\begin{description}
\item[A]: $|m|\geq m_{\rm eq}(\beta\phi_{\rm max})$,
\item[B]: $m_{\rm sp}(\beta\phi_{\rm max})\leq |m|<m_{\rm eq}(\beta\phi_{\rm max})$,
\item[C]: $|m|<m_{\rm sp}(\beta\phi_{\rm max})$.
\end{description}
Compare it with the parameter space decomposition in the van der Waals limit.
The difference is appearing $\beta\phi_{\rm max}$ instead of $\beta$.
The ``phase diagram'' on the $(\beta,m)$-plane for $\phi_{\rm max}=0.31$, 
which corresponds to $\phi(\bi{x})=1/x$ in two dimensions, is given in Fig.~\ref{fig:phase_TM}.

\subsection{Exactness of MF theory in the canonical ensemble}
\label{sec:exactness_can}

In this section, we assume C2.
The free energy in the canonical ensemble $g(\beta,h)$ is determined by finding the minimum of $f(\beta,m)-h\cdot m$, see Eq.~(\ref{eq:g_from_f}).
The minimum point, which corresponds to the MF magnetization curve on $(\beta,m)$ plane,
 is always in the region A (see Fig.~\ref{fig:phase_TM}), and hence exactness of MF theory is shown for the canonical ensemble:
\begin{corollary}[Exactness of MF theory in the canonical ensemble]
We consider the canonical ensemble associated with the Hamiltonian~(\ref{eq:Hamiltonian}) 
on the $d$-dimensional regular lattice with the periodic boundary condition.
If the interaction potential, which is normalized as $\int_{\Omega_1^d}d^dx\phi(\bi{x})=1$, satisfies C1 and C2, then
\beq
g(\beta,h)=g_{\rm MF}(\beta,h)
\eeq
for any $\beta>0$ and any $h$ in the non-additive limit.
\end{corollary}
{\it Proof}.
From Theorem~\ref{theorem:main}, the inequality
\beq
f_{\rm MF}^{**}(\beta,m)-\phi_{\rm max}\Delta f_{\rm MF}(\beta\phi_{\rm max},m)\leq f(\beta,m)\leq f_{\rm MF}(\beta,m)
\eeq
always holds.
As we will see during the proof of Theorem~\ref{theorem:main} in Sec.~\ref{sec:proof}, the LHS is a decreasing function of $\phi_{\rm max}$.
Because $\phi_{\rm max}\leq 1$, the inequality
\beq
f_{\rm MF}^{**}(\beta,m)\leq f(\beta,m)\leq f_{\rm MF}(\beta,m)
\eeq
also holds.
It yields
\beq
\inf_m\left[f_{\rm MF}^{**}(\beta,m)-h\cdot m\right]\leq g(\beta,h)\leq\inf_m\left[f_{\rm MF}(\beta,m)-h\cdot m\right].
\eeq
From the definition, $\inf_m[f_{\rm MF}(\beta,m)-h\cdot m]=g_{\rm MF}(\beta,h)$.
Since the convex envelope satisfies Eq.~(\ref{eq:f_from_g}), whose proof is given in \ref{app:convex_envelope}, we have
\begin{eqnarray}
\inf_m\left[f_{\rm MF}^{**}(\beta,m)-h\cdot m\right]
&=\inf_m\sup_{h'}\left[g_{\rm MF}(\beta,h')+(h'-h)\cdot m\right]
\nonumber \\
&\geq\inf_mg_{\rm MF}(\beta,h)=g_{\rm MF}(\beta,h),
\end{eqnarray}
where we put $h'=h$ in the second line.
Thus we have $g(\beta,h)=g_{\rm MF}(\beta,h)$.

\subsection*{Remark on the random spin systems}
Let us consider a random spin system whose Hamiltonian is given by
\beq
H=-\frac{1}{2}\sum_{i,j}^N\xi_{ij}J_{ij}\sigma_i\sigma_j,
\eeq
where $\{\xi_{ij}\}$ are the identical and independent Gaussian random variables of mean zero and variance one.
By applying the replica method~\cite{Nishimori_text}, we can show that the MF theory is exact in the canonical ensemble if $J_{ij}^2=L^{-d}\phi(\bi{r}_{ij}/L)$,
where $\phi(\bi{x})$ satisfies C1 and C2~\cite{Mori2011_instability}.
For example, if we consider the case of $J_{ij}\sim 1/r_{ij}^{\alpha}$, the MF theory is exact when $\alpha<d/2$.
Strictly speaking, the result is not rigorous as the replica method is used in the proof,
but the exactness of MF theory was confirmed numerically in Ref.~\cite{Wittmann-Young2012}.

\subsection{Appearance of non-MF phase in the restricted canonical ensemble}
\label{sec:non-MF}

A major difference from the case of the van der Waals limit is stability of a state in the region B.
In the region B for the van der Waals limit, a homogeneous state is always globally unstable.
However, in the region B for the non-additive limit, a homogeneous state might be globally stable in some cases,
and in those cases the MF theory is still exact in the region B.

The phase transition occurs somewhere in the region B.
In the {\it MF phase}, $f(\beta,m)=f_{\rm MF}(\beta,m)$ and in the {\it non-MF phase}, $f(\beta,m)<f_{\rm MF}(\beta,m)$.
A typical equilibrium configuration is inhomogeneous in the non-MF phase; see Ref.~\cite{Mori2010_analysis}.
Thus, we could rigorously show the existence of the phase transition, 
where the spacial structure appears due to a non-constant and non-additive interaction $\phi(\bi{x})$.
Such a phase transition cannot be captured by the MF models.

\subsection{Brief sketch of proof of Theorem~\ref{theorem:main}}
\label{sec:proof}

The method of coarse graining plays an important role to prove Theorem\ref{theorem:main}.
The idea of this method was developed by van Kampen~\cite{van_Kampen1964} and made rigorous by Lebowitz and Penrose~\cite{Lebowitz-Penrose1966}
in the analysis of the van der Waals limit.
It has been shown that the method of coarse graining is naturally extended 
to the case of the non-additive limit~\cite{Barre2005,Mori2010_analysis,Mori2011_instability}.

We divide the system into many cells, each of which is a hypercube of side $l$.
Each cell is denoted by ${\rm C}_p$, $p=1,2,\dots,(L/l)^2$.
The coarse graining is a procedure to replace every spin $\sigma_i$ in a cell ${\rm C}_p$ by the average magnetization of this cell, i.e.,
\beq
\sigma_i\rightarrow\frac{1}{l^d}\sum_{j\in{\rm C}_p}\sigma_j=:m_p
\eeq
for all $i\in{\rm C}_p$.
By this procedure, the Hamiltonian becomes
\begin{eqnarray} \fl
H_h\rightarrow \tilde{H}_h&:=-\frac{l^{2d}}{2}\sum_{p,q}^{(L/l)^d}\gamma^d
\left(\frac{1}{l^{2d}}\sum_{i\in{\rm C}_p}\sum_{j\in{\rm C}_q}\phi(\gamma\bi{r}_{ij})\right)m_pm_q
-l^dh\cdot\sum_{p=1}^{(L/l)^d}m_p
\nonumber \\ \fl
&=-\frac{l^{2d}}{2}\sum_{p,q}^{(L/l)^d}\gamma^d\phi_{pq}m_pm_q-l^dh\cdot\sum_{p=1}^{(L/l)^d}m_p,
\end{eqnarray}
where we have defined the coarse-grained interaction by
\beq
\phi_{pq}:=\frac{1}{l^{2d}}\sum_{i\in{\rm C}_p}\sum_{j\in{\rm C}_q}\phi(\gamma\bi{r}_{ij}).
\eeq
The new Hamiltonian $\tilde{H}_h$ is called the coarse-grained Hamiltonian.

We take either of the non-additive limit or the van der Waals limit.
We then take the limit of $l\rightarrow\infty$ finally.
It means that the size of each cell is very large compared to microscopic length scale
but much smaller than the size of the whole system and the interaction range.

The goal is to compute the free energy density in the restricted canonical ensemble,
\beq
f(\beta,m)=-{\rm Lim}\frac{1}{L^d\beta}\ln\sum_{\{\sigma_i\in{\cal S}\}}\chi\left(\sum_{i=1}^N\sigma_i=Nm\right)e^{-\beta H_0}.
\eeq
The coarse-grained free energy is defined in a similar way:
\beq
\tilde{f}(\beta,m):=-\lim_{l\rightarrow\infty}{\rm Lim}\frac{1}{L^d\beta}
\ln\sum_{\{\sigma_i\in{\cal S}\}}\chi\left(\sum_{i=1}^N\sigma_i=Nm\right)e^{-\beta \tilde{H}_0}.
\eeq
We can show that
\beq
f(\beta,m)=\tilde{f}(\beta,m),
\label{eq:coarse-graining}
\eeq
i.e., the coarse graining can be done without any approximation in a long-range interacting system.
Proof of Eq.~(\ref{eq:coarse-graining}) is seen in Ref.~\cite{Mori2011_instability}.

If we admit Eq.~(\ref{eq:coarse-graining}), it is not so difficult to derive the parameter space decomposition.
After the above mentioned limiting procedure ($l\rightarrow\infty$ is taken after the limit ``Lim''),
the coarse-grained free energy is expressed by the following variational form:
\beq
\tilde{f}(\beta,m)={\rm Lim}\frac{1}{L^d}\min_{\{ m_p\}:\frac{1}{l^d}\sum_{p=1}^{(L/l)^d}m_p=m}
\left[\tilde{H}_0-\frac{l^d}{\beta}\sum_{p=1}^{(L/l)^d}s(m_p)\right].
\label{eq:cg_free_discrete}
\eeq
In the non-additive limit, we can express the coarse-grained free energy in the continuum form:
\beq
\tilde{f}(\beta,m)=\min_{\{ m(\bi{x})\}:\int_{\Omega_1^d}d^dxm(\bi{x})=m}{\cal F}(\beta,\{ m(\bi{x})\}).
\label{eq:cg_free}
\eeq
Here ${\cal F}(\beta,\{ m(\bi{x})\})$ denotes the {\it free energy functional},
\beq \fl
{\cal F}(\beta,\{ m(\bi{x})\}):=-\frac{1}{2}\int_{\Omega_1^d\times\Omega_1^d}d^dxd^dyU(\bi{x}-\bi{y})m(\bi{x})m(\bi{y})
-\frac{1}{\beta}\int_{\Omega_1^d}d^dxs(m(\bi{x})).
\eeq
The scaled potential $U(\bi{x})$ is defined as
\beq
U(\bi{x}):={\rm Lim}(\gamma L)^d\phi(\gamma L\bi{x}).
\eeq
In the non-additive limit, $\gamma L=1$ and thus $U(\bi{x})$ is identical to $\phi(\bi{x})$.

In the van der Waals limit, the formula~(\ref{eq:cg_free_discrete}) is correct if the potential $\phi(\bi{x})$ satisfies C1.
On the other hand, the use of the continuum form~(\ref{eq:cg_free}) requires C1 and C2.
In that case, 
${\rm Lim}=\lim_{\gamma\rightarrow 0}\lim_{L\rightarrow\infty}$ and the normalization $\int_{{\mathbb R}^d}d^dx\phi(\bi{x})=1$
yields $U(\bi{x})=\delta(\bi{x})$.
Thus,  when the interaction potential satisfies C1 and C2, the van der Waals limit is equivalent to formally putting $\phi(\bi{x})=\delta(\bi{x})$ in the non-additive limit
as far as thermodynamic properties are concerned.
We are now interested in the non-additive limit, and hence we simply put $U(\bi{x})=\phi(\bi{x})$ in later calculations.
If one wants to consider the van der Waals limit, one just has to put $\phi(\bi{x})=\delta(\bi{x})$.

An upper bound of the free energy is easily obtained by putting $m(\bi{x})=m$ in Eq.~(\ref{eq:cg_free}):
\beq 
f(\beta,m)\leq {\cal F}(\beta,\{ m\})=-\frac{1}{2}m^2-\frac{1}{\beta}s(m)=f_{\rm MF}(\beta,m).
\label{eq:upper}
\eeq
Thus the MF theory is exact as long as we consider only the homogeneous spin configurations.
The presence of inequality implies that there might be some inhomogeneous spin configurations which has a lower free energy.

Up to now, the boundary condition does not matter.
Here we impose the periodic boundary condition.
The Fourier transformation diagonalizes the coarse-grained Hamiltonian:
\beq
-\frac{1}{2}\int_{\Omega_1^d\times\Omega_1^d}d^dxd^dy\phi(\bi{x}-\bi{y})m(\bi{x})m(\bi{y})
=-\frac{1}{2}\sum_{\bi{n}\in{\mathbb Z}^d}\phi_{\bi{n}}|m_{\bi{n}}|^2,
\eeq
where the Fourier component is defined as
\beq
m_{\bi{n}}:=\int_{\Omega_1^d}d^dxe^{2\pi i\bi{n}\cdot\bi{x}}m(\bi{x}).
\eeq

From the normalization $\phi_0=1$,
\beq
-\frac{1}{2}\sum_{\bi{n}\in{\mathbb Z}^d}\phi_{\bi{n}}|m_{\bi{n}}|^2
=-\frac{1}{2}m^2-\frac{1}{2}\sum_{\bi{n}\in{\mathbb Z}^d\backslash 0}\phi_{\bi{n}}|m_{\bi{n}}|^2.
\eeq
Because $\phi_{\rm max}$ is defined by Eq.~(\ref{eq:phi_max}), we have
\begin{eqnarray} \fl
-\frac{1}{2}\sum_{\bi{n}\in{\mathbb Z}^d}\phi_{\bi{n}}|m_{\bi{n}}|^2
&\geq -\frac{1}{2}m^2-\frac{1}{2}\phi_{\rm max}\sum_{\bi{n}\in{\mathbb Z}^d}|m_{\bi{n}}|^2
\label{eq:monotone_phi}
\\ \fl
&=-\frac{1}{2}m^2+\frac{1}{2}\phi_{\rm max}m^2-\frac{1}{2}\phi_{\rm max}\int_{\Omega_1^d}d^dx m(\bi{x})^2.
\end{eqnarray}
Note that the RHS of Eq.~(\ref{eq:monotone_phi}) is obviously a monotonically non-increasing function of $\phi_{\rm max}$.
The above inequality yields
\begin{eqnarray}\fl
{\cal F}(\beta,\{ m(\bi{x})\})
&\geq -\frac{1}{2}m^2+\frac{1}{2}\phi_{\rm max}m^2
+\phi_{\rm max}\int_{\Omega_1^d}d^dx\left(-\frac{1}{2}m(\bi{x})^2-\frac{1}{\beta\phi_{\rm max}}s(m(\bi{x}))\right)
\nonumber \\ \fl
&= \left(-\frac{1}{2}m^2-\frac{1}{\beta}s(m)\right)-\phi_{\rm max}\left(-\frac{1}{2}m^2-\frac{1}{\beta\phi_{\rm max}}s(m)\right)
\nonumber \\ \fl
&+\phi_{\rm max}\int_{\Omega_1^d}d^dx\left(-\frac{1}{2}m(\bi{x})^2-\frac{1}{\beta\phi_{\rm max}}s(m(\bi{x}))\right).
\end{eqnarray}
Since the MF free energy is given by $f_{\rm MF}(\beta,m)=-m^2/2-(1/\beta)s(m)$, the lower bound is written only in terms of the MF free energy,
\begin{eqnarray}\fl
{\cal F}(\beta,\{ m(\bi{x})\})\geq &f_{\rm MF}(\beta,m)
\nonumber \\ \fl
&-\phi_{\rm max}\left[f_{\rm MF}(\beta\phi_{\rm max},m)-\int_{\Omega_1^d}d^dx f_{\rm MF}(\beta\phi_{\rm max},m(\bi{x}))\right].
\end{eqnarray}
We can show that~\cite{Mori2011_instability}
\beq
\min_{\{ m(\bi{x})\}:\int_{\Omega_1^d}d^dx m(\bi{x})=m}\int_{\Omega_1^d}d^dx f_{\rm MF}(\beta,m(\bi{x}))=f_{\rm MF}^{**}(\beta,m).
\eeq
By using this formula and from Eqs.~(\ref{eq:coarse-graining}) and (\ref{eq:cg_free}), we obtain
\beq
f(\beta,m)\geq f_{\rm MF}(\beta,m)-\phi_{\rm max}\Delta f_{\rm MF}(\beta\phi_{\rm max},m).
\label{eq:lower}
\eeq
Because the RHS of Eq.~(\ref{eq:monotone_phi}) is a monotonically non-increasing function of $\phi_{\rm max}$,
the above lower bound is also a monotonically non-increasing function of $\phi_{\rm max}$.

Combining the lower bound~(\ref{eq:lower}) with the upper bound~(\ref{eq:upper}), we obtain
\beq
f_{\rm MF}(\beta,m)-\phi_{\rm max}\Delta f_{\rm MF}(\beta\phi_{\rm max},m)\leq f_{\rm MF}(\beta,m)\leq f_{\rm MF}(\beta,m).
\label{eq:inequality}
\eeq
From the above inequality, if $f_{\rm MF}(\beta\phi_{\rm max},m)=f_{\rm MF}^{**}(\beta\phi_{\rm max},m)$
(or $\Delta f_{\rm MF}(\beta\phi_{\rm max},m)=0$),
the LHS and the RHS coincide and thus $f(\beta,m)=f_{\rm MF}(\beta,m)$.
This is the region A.

The decomposition of the remaining part into the region B and the region C is based on the standard linear stability analysis around the uniform spin configuration.
If the free energy functional increases for any infinitesimal displacement of the magnetization field, $m(\bi{x})=m+\delta m(\bi{x})$
with $\int_{\Omega_1^d}d^dx\delta m(\bi{x})=0$, the uniform state is said to be locally stable. Otherwise it is locally unstable.

Stability is determined by the lowest eigenvalue of the Hessian matrix whose $((\bi{x}_1,a),(\bi{x}_2,b))$
$\left.\delta^2{\cal F}(\beta,\{ m(\bi{x})\})/\delta m^{(a)}(\bi{x}_1)\delta m^{(b)}(\bi{x}_2)\right|_{m(\bi{x})=m}.$
In terms of Fourier components,
\begin{eqnarray}
\left.\frac{\delta^2{\cal F}(\beta,\{ m(\bi{x})\})}{\delta m_{\bi{n}}^{(a)}\delta m_{\bi{l}}^{(b)}}\right|_{m(\bi{x})=m}
&=\delta_{\bi{n}+\bi{l},0}\phi_{\bi{n}}\frac{\d^2f_{\rm MF}(\beta\phi_{\bi{n}},m)}{\d m^{(a)}\d m^{(b)}}
\nonumber \\
&=\delta_{\bi{n}+\bi{l},0}\phi_{\bi{n}}{\sf H}_{ab}(\beta\phi_{\bi{n}},m).
\end{eqnarray}
Since the lowest eigenvalue of the matrix ${\sf H}(\beta,m)$ is a non-increasing function of $\beta$,\footnote
{This fact is resulted from the concavity of the configurational entropy $s(m)$.}
the sign of the lowest eigenvalue of the Hessian matrix coincides with the sign of the lowest eigenvalue of ${\sf H}(\beta\phi_{\rm max},m)$,
that is $\lambda(\beta\phi_{\rm max},m)$ by definition.

In this way, if $\lambda(\beta\phi_{\rm max})\geq 0$, the state with a uniform spin configuration is at least locally stable (or marginal)
and the result of the MF model may or may not be exact.
This is the region B.

If $\lambda(\beta\phi_{\rm max},m)<0$, the state with a uniform spin configuration is unstable and the second equality of Eq.~(\ref{eq:inequality}) is not realized.
This is the region C.

\section{Without non-negativity of the interaction potential}
\label{sec:non-negativity}

The condition of non-negativity C2 ensures $\phi_{\rm max}\leq 1$, 
which was essential for exactness of MF theory in the canonical ensemble given in Sec.~\ref{sec:exactness_can}.
In other words, all the points on the $(\beta,h)$ plane belong to the region A when $\phi_{\rm max}\leq 1$ is satisfied.
Even if the condition C2 is not satisfied, the above conclusion immediately follows from the parameter space decomposition
as long as $\phi_{\rm max}\leq 1$.

The situation becomes different when $\phi_{\rm max}>1$.
In this case, the fictitious temperature $(\beta\phi_{\rm max})^{-1}$ is lower than the genuine temperature $\beta^{-1}$.
We cannot conclude that a given point on the $(\beta,h)$-plane is always in the region A;
the MF theory might not be exact even in the canonical ensemble.

\begin{figure}[t]
\begin{center}
\includegraphics[clip,width=8cm]{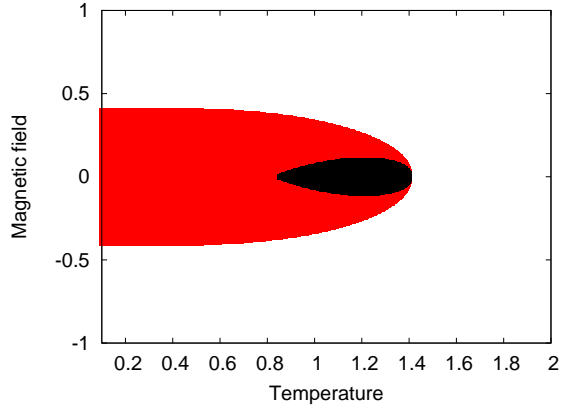}
\caption{The phase diagram of the long-range interacting Ising model with $\phi_{\rm max}=1.40$.
The white, red (gray), and black regions correspond to the region A, B, and C.}
\label{fig:phase_TH}
\end{center}
\end{figure}

Again let us consider the Ising variable ${\cal S}=\{\pm 1\}$ as an example.
We consider the canonical ensemble.
The equilibrium magnetization in the MF model is denoted by $m_{\rm eq}(\beta,h)$.\footnote
{$m_{\rm eq}(\beta)$ defined in Sec.~\ref{sec:MF} is identical with $m_{\rm eq}(\beta,0)$.}
It is also expresses as $m_{\rm eq}={\rm arginf}_m[f_{\rm MF}(\beta,m)-h\cdot m]$.
From the parameter space decomposition, if $(\beta,m_{\rm eq}(\beta,h))$ on the $(\beta,m)$-plane is in the region A, the MF theory is exact,
but if $(\beta,m_{\rm eq}(\beta,h))$ is in the region C, the MF theory is not exact.
If $(\beta,m_{\rm eq}(\beta,m))$ is in the region B, the uniform state predicted by the MF model is at least locally stable.
Thus we can obtain the ``phase diagram'' on the $(\beta,h)$ plane, which is depicted in Fig.~\ref{fig:phase_TH}. 

\begin{figure}[t]
\begin{center}
\begin{tabular}{cc}
\includegraphics[clip,width=6cm]{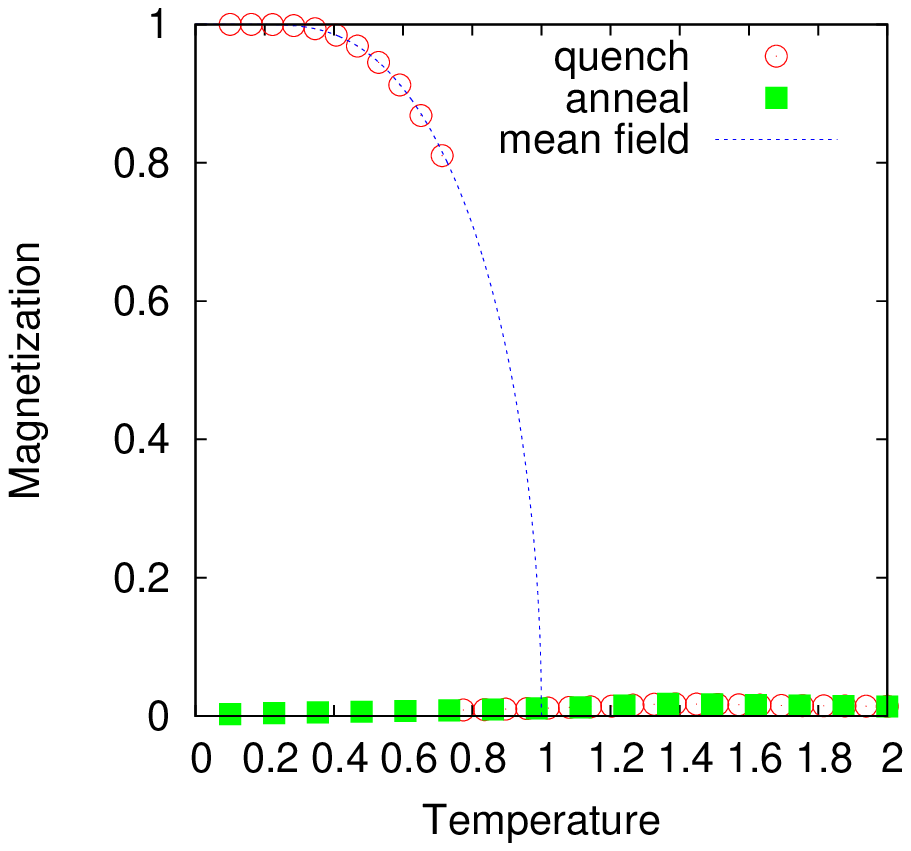}&
\includegraphics[clip,width=6cm]{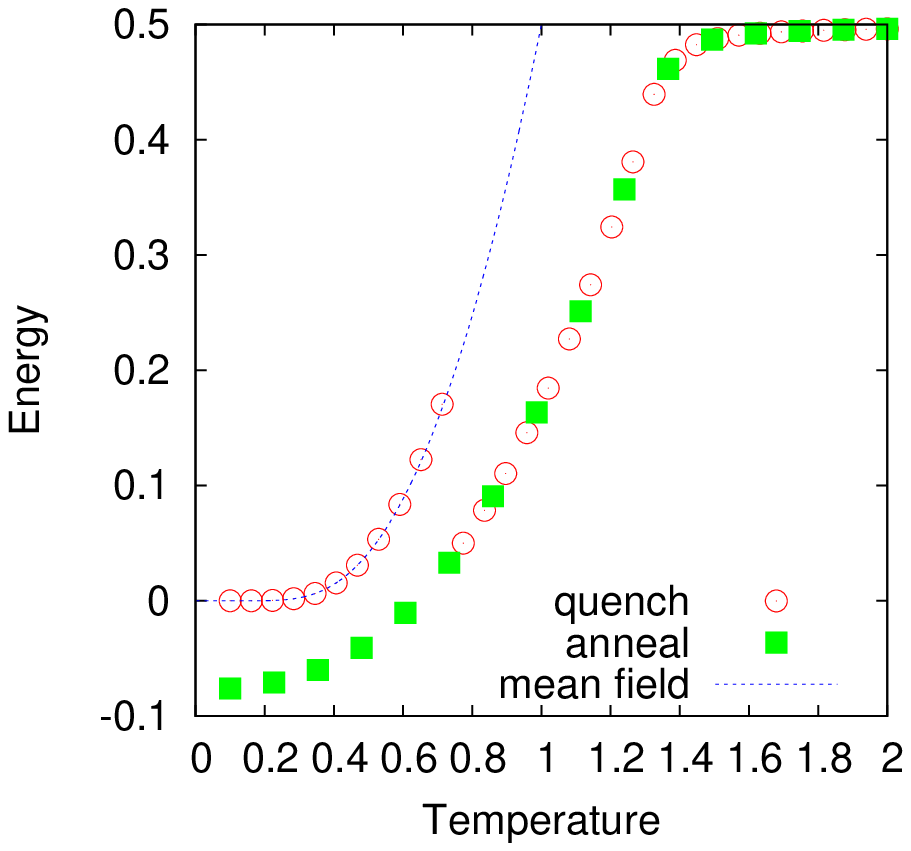}
\end{tabular}
\caption{(Left) The absolute magnetization density $|m|$ is plotted against the temperature $T=\beta^{-1}$.
(Right) The energy density is plotted against the temperature $T=\beta^{-1}$.
The zero point of the energy is chosen so that it becomes zero in the all up (or down) state.}
\label{fig:equil}
\end{center}
\end{figure}

\begin{figure}[t]
\begin{center}
\begin{tabular}{cc}
(a) &(b) \\
\includegraphics[clip,width=5cm]{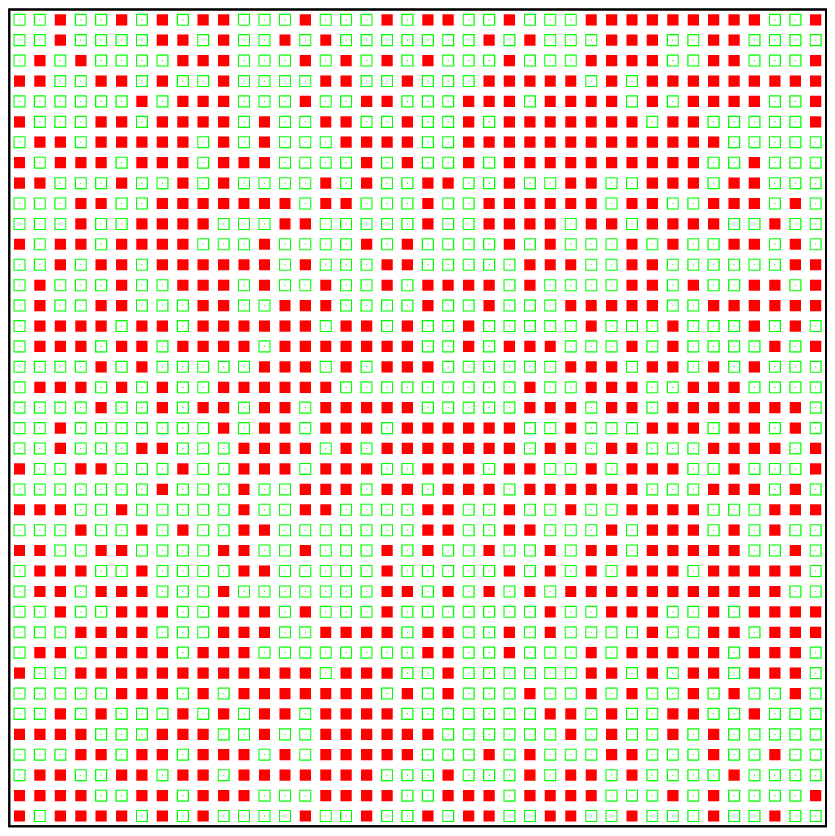}&
\includegraphics[clip,width=5cm]{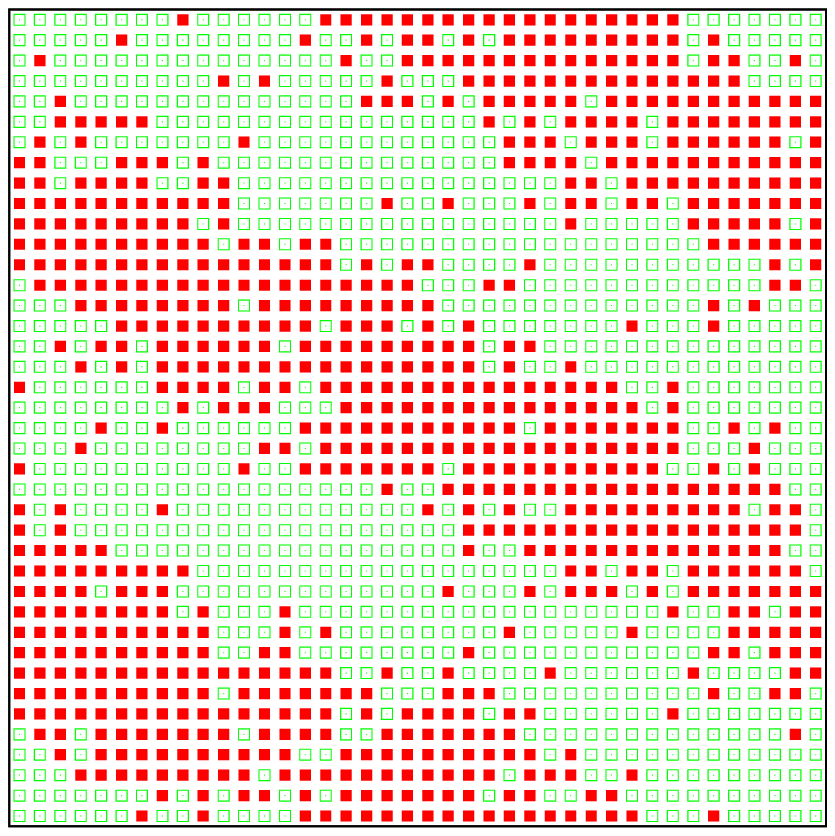}\\
(c)&(d) \\
\includegraphics[clip,width=5cm]{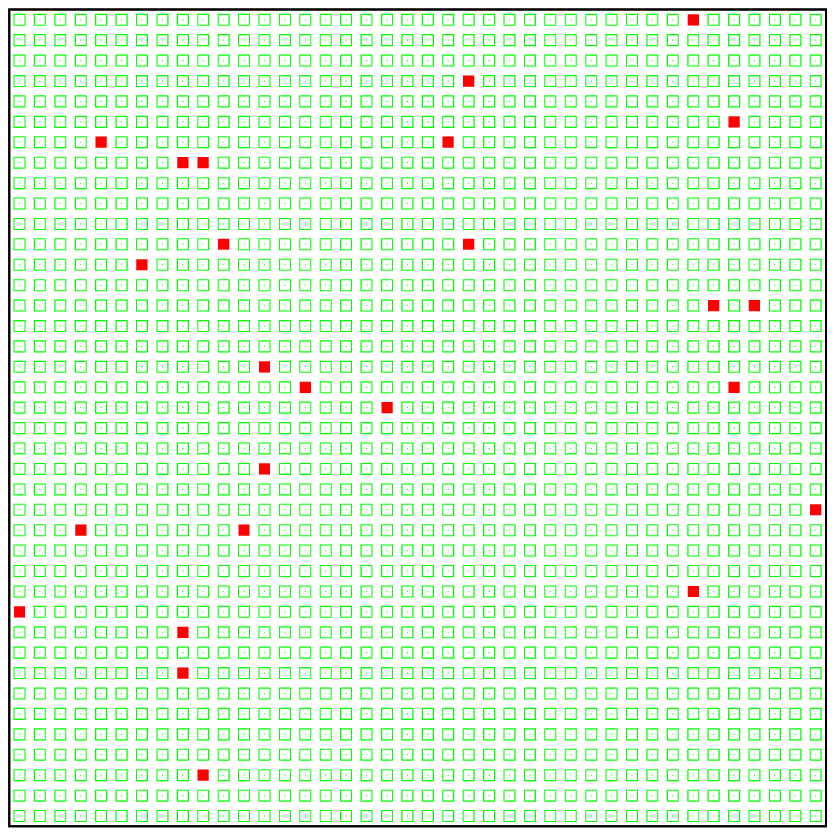}&
\includegraphics[clip,width=5cm]{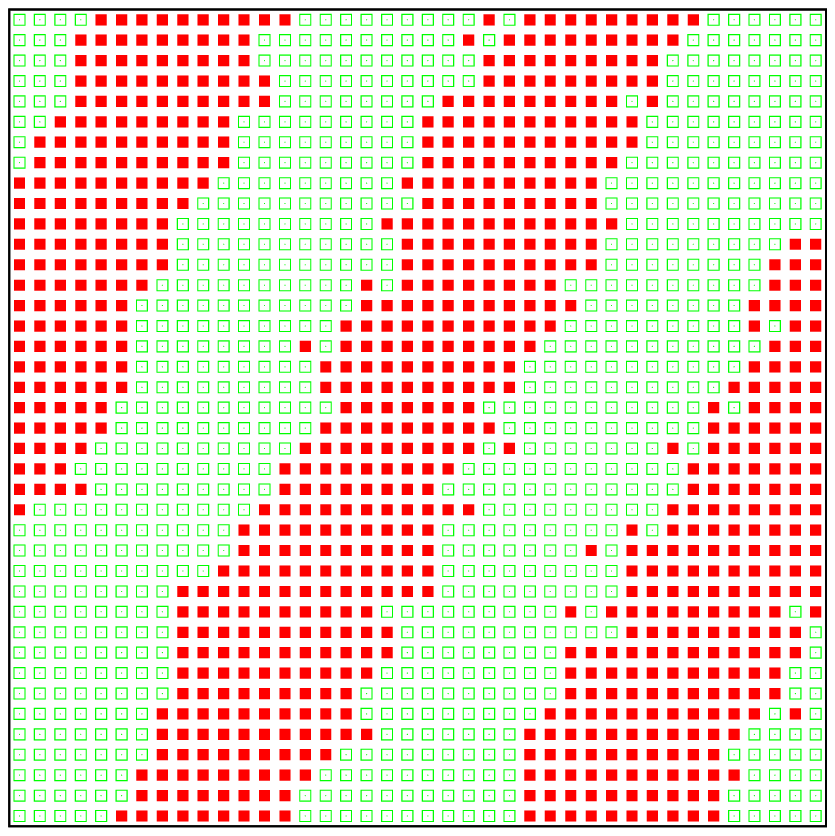}
\end{tabular}
\caption{Typical equilibrium configurations for (a) $(\beta^{-1},h)=(1.6,0)$, which belongs to the region A, 
(b) $(\beta^{-1},h)=(1.0,0)$, which belongs to the region C, 
and (c,d) $(\beta^{-1},h)=(0.48,0)$ in the quench and anneal dynamics, respectively, which belong to the region B.
Red full squares and green empty squares correspond to the site with $\sigma_i=+1$ and $-1$, respectively.
The system size is $L=40$.}
\label{fig:configuration}
\end{center}
\end{figure}

The Monte-Carlo simulation is performed for the Ising model with a quite artificial interaction $\phi(\bi{x})\propto [1+8\cos(4\pi x)]/x$ 
on the two-dimensional square lattice of side $L=80$,
and the equilibrium magnetization density and the energy density are plotted as a function of the temperature in Fig.~\ref{fig:equil}.
The value of $\phi_{\rm max}$ of this interaction potential is about 1.40 and the phase diagram is identical to that in Fig.~\ref{fig:phase_TH}.
We present the result when no magnetic field is applied, $h=0$.
The red circles correspond to the numerical results when the initial state is all up state, i.e., $\sigma_i=+1$ $\forall i$,
and a thermal bath at temperature $\beta^{-1}$ is suddenly attached.
The dynamics generated by this procedure is called ``quench dynamics''.
The green squares are numerical results when the temperature of the thermal bath is slowly decreased from above.
We call the dynamics generated by this procedure ``anneal dynamics''.
Above $T\simeq 1.4$, the system is in the region A and equilibrium configurations are uniform; see Fig.~\ref{fig:configuration} (a).
Below $T\simeq 1.4$ the system enters the region C and equilibrium configurations become inhomogeneous; see Fig.~\ref{fig:configuration} (b).
Below $T\simeq 0.8$, the system enters the region B and bistable branch appears; the red circles and green squares do not coincide.
In this low-temperature region, the red circles correspond to homogeneous states (see Fig.~\ref{fig:configuration} (c)), 
which are identical to equilibrium states predicted by the MF model,
and the green squares correspond to inhomogeneous states not predicted by the MF model; see Fig.~\ref{fig:configuration} (d).
As you can see in Fig.~\ref{fig:phase_TH} (b), the green squares have lower energies than the red circles at low temperatures.
Therefore, in this model, uniform states predicted by the MF model are just metastable states
and the genuine equilibrium states are inhomogeneous below $T\simeq 0.8$.

Typical spin configurations are shown in Fig.~\ref{fig:configuration}.
When the parameter belongs to the region A in Fig.~\ref{fig:phase_TH}, an equilibrium state is homogeneous,
while when the parameter belongs to the region C, an equilibrium state is inhomogeneous, which reflects the specific shape of the interaction potential.

\subsection*{Remark on the van der Waals limit}
In Sec.~\ref{sec:decomposition}, it has been mentioned that when the interaction potential satisfies C1 and C2,
the van der Waals limit is recovered by formally putting $\phi(\bi{x})=\delta(\bi{x})$ in the analysis for the non-additive limit.
If we put $\phi(\bi{x})=\delta(\bi{x})$ in Eq.~(\ref{eq:cg_free}), we can easily show that $f(\beta,m)=f_{\rm MF}^{**}(\beta,m)$,
which is the Lebowitz-Penrose theorem~\cite{Lebowitz-Penrose1966}.

However, if C2 does not hold, the van der Waals limit cannot be recovered by putting $\phi(\bi{x})=\delta(\bi{x})$.
In that case we cannot use the continuum expression, Eq.~(\ref{eq:cg_free}).
We have to analyze in the discrete form, Eq.~(\ref{eq:cg_free_discrete}).
In the van der Waals limit, we define $\phi_{\bi{n}; L,\gamma}$ as
\beq
\phi_{\bi{n};L,\gamma}:=\int_{\Omega_1^d}d^dx\phi(\gamma L\bi{x})e^{2\pi i\bi{n}\cdot\bi{x}}.
\eeq
We define $\phi_{{\rm max}; L,\gamma}:=\max_{\bi{n}\in\mathbb{Z}^d}\phi_{\bi{n};L,\gamma}$
and
\beq
\phi_{\rm max}^{\rm vdW}:=\lim_{\gamma\rightarrow 0}\lim_{L\rightarrow\infty}\phi_{{\rm max};L,\gamma}.
\eeq

If $\phi_{\rm max}^{\rm vdW}=1$, then we can justify the use of the continuum expression~(\ref{eq:cg_free}) and we can show the Lebowitz-Penrose theorem.
However, when C2 does not hold, $\phi_{\rm max}^{\rm vdW}=1$ is not satisfied in general.
When $\phi_{\rm max}^{\rm vdW}>1$, from Eq.~(\ref{eq:cg_free_discrete}), we can show that
\beq
f_{\rm MF}(\beta,m)-\phi_{\rm max}\Delta f_{\rm MF}(\beta\phi_{\rm max},m)\leq f(\beta,m)\leq f_{\rm MF}^{**}(\beta,m).
\eeq
Gates and Penrose~\cite{Gates-Penrose1969,Gates-Penrose1970-1,Gates-Penrose1970-2} 
obtained more precise upper bound and showed that there is a certain parameter region 
where the free energy is strictly less than the convex envelope of the MF free energy, i.e., $f(\beta,m)<f_{\rm MF}^{**}(\beta,m)$
when $\phi_{\rm max}^{\rm vdW}>2$.
In this parameter region, the inhomogeneity appears at a length scale about $\gamma^{-1}$.

\section{Conclusion}
\label{sec:conclusion}
In this paper, the concepts of additivity and extensivity have been reconsidered, and the consequence of non-additivity has been reviewed.
The relation between stability of a uniform state and the convexity of the free energy has been extended to the non-additive regime (Theorem~\ref{theorem:main}).
Based on Theorem~\ref{theorem:main}, the exactness of MF theory has been discussed.
In particular, it has been found that if the interaction does not satisfy the non-negativity,
the system may undergo the phase transition between the MF phase and the non-MF phase even in the canonical ensemble.

Statistical mechanics of long-range interacting systems will be relevant for several situations even if the interaction is not macroscopically long.
In a relatively small system, it is not additive if the interaction range is comparable with the system size.
There have been several theoretical and experimental attempts to realize a long-range interacting system in laboratory~\cite{ODell2000,Chalony2013}.
Moreover, recently, it has been argued that a certain short-range interacting system exhibits non-additivity in {\it quasi}-equilibrium states,
and those quasi-equilibrium states are described by equilibrium statistical mechanics of a long-range interacting system~\cite{Mori2013_nonadditivity}.
In this way, statistical mechanics of non-additive systems will help us to study such situations
and its range of applicability would be broader than we have expected.

\section*{Acknowledgments}
The author thanks Seiji Miyashita for continual discussion on this and related subjects.
This work was supported by JSPS (Grant No. 227835) and the Sumitomo Foundation.

\appendix
\section{Proof of Eq.~(\ref{eq:f_from_g})}
\label{app:convex_envelope}
By definition, $g_{\rm MF}(\beta,h)=\inf_{m'}[f_{\rm MF}(\beta,m')-h\cdot m']$ and
\beq
\sup_h\left[g_{\rm MF}(\beta,h)+h\cdot m\right]
=\sup_h\inf_{m'}\left[f_{\rm MF}(\beta,m')+h\cdot(m-m')\right].
\eeq
We obtain an upper bound by putting $m'=m$:
\beq
\sup_h\left[g_{\rm MF}(\beta,h)+h\cdot m\right]\leq f_{\rm MF}(\beta,m).
\eeq
On the other hand,
\beq \fl
\sup_h\inf_{m'}\left[f_{\rm MF}(\beta,m')+h\cdot(m-m')\right]
\geq\sup_h\inf_{m'}\left[f_{\rm MF}^{**}(\beta,m')+h\cdot(m-m')\right].
\eeq
Since $f_{\rm MF}^{**}(\beta,m')$ is a convex function of $m'$, for an arbitrary fixed $m$, 
there exists $h'\in{\mathbb R}$ (the ``subgradient'' of $f_{\rm MF}^{**}$) such that
\beq
f_{\rm MF}^{**}(\beta,m')\geq f_{\rm MF}^{**}(\beta,m)+h'\cdot(m'-m)
\eeq
for all $m'$.
By using this property, we obtain a lower bound:
\beq \fl
\sup_h\left[g_{\rm MF}(\beta,h)+h\cdot m\right]\geq\sup_h\inf_{m'}\left[f_{\rm MF}^{**}(\beta,m)+(h-h')\cdot(m-m')\right]
\geq f_{\rm MF}^{**}(\beta,m).
\eeq
By collecting upper and lower bounds,
\beq
f_{\rm MF}^{**}(\beta,m)\leq\sup_h\left[g_{\rm MF}(\beta,h)+h\cdot m\right]\leq f_{\rm MF}(\beta,m).
\label{eq:app_inequality}
\eeq
From the observation that
\begin{eqnarray}
\lambda\sup_h\left[g_{\rm MF}(\beta,h)+h\cdot m_1\right]+(1-\lambda)\sup_h\left[g_{\rm MF}(\beta,h)+h\cdot m_2\right]
\nonumber \\
\geq\sup_h\left[g_{\rm MF}(\beta,h)+h\cdot (\lambda m_1+(1-\lambda)m_2)\right]
\end{eqnarray}
for any $0\leq\lambda\leq 1$, obviously $\sup_h\left[g_{\rm MF}(\beta,h)+h\cdot m\right]$ is a convex function of $m$.
Because $f_{\rm MF}^{**}$ is the maximum convex function not exceeding $f_{\rm MF}$,
the convex function satisfying the inequality~(\ref{eq:app_inequality}) is only
\beq
\sup_h\left[g_{\rm MF}(\beta,h)+h\cdot m\right]=f_{\rm MF}^{**}(\beta,m).
\eeq

\section*{Reference}
%\bibliography{LR_ref.bib}

\end{document}